\documentclass[9pt]{entcs}
\usepackage{entcsmacro}

\usepackage{rotating}

\usepackage{float}
\usepackage[utf8]{inputenc}
\usepackage{verbatim}
\usepackage{adjustbox}
\usepackage{marvosym}
\usepackage{graphicx}
\usepackage{amsmath}
\usepackage{amscd}
\usepackage{xcolor}
\usepackage{array}

\usepackage{bbold}
\usepackage{url}
\usepackage{upgreek}
\usepackage{stmaryrd}
\usepackage{enumitem}

\usepackage{lipsum}
\usepackage{tikz-cd}
\usetikzlibrary{cd}
\usetikzlibrary{calc}
\usetikzlibrary{arrows}

\usepackage{bussproofs}
\EnableBpAbbreviations

\usepackage{hyperref}

\usepackage{pdflscape}

\usepackage{listings}
\DeclareMathAlphabet{\mathpzc}{OT1}{pzc}{m}{it}

\newcommand{\id}{id}

\newcommand{\Set}{\mathsf{Set}}

\newcommand{\ar}[1]{\##1}

\newcommand{\ol}[1]{\overline{#1}}

\newcommand{\funccat}[2]{[#1\mathbin{,}#2]}%
\newcommand{\type}[1]{\mathsf{#1}}%

\newcolumntype{P}[1]{>{\centering\arraybackslash}p{#1}}
\newcommand*{\thead}[1]{\multicolumn{1}{|c|}{\bfseries \emph{#1}}}

\newcommand{\set}{\mathsf{Set}}

\renewcommand{\id}{\mathit{id}}

\newcommand{\lop}{\langle}
\newcommand{\rop}{\rangle}

\newcommand{\len}[1]{\lvert #1 \rvert}

\def\lastname{Johann and Cagne}

\begin{document}
\begin{frontmatter}
\vspace*{-0.075in}
\title{How Functorial Are 
    (Deep) GADTs?\vspace*{-0.2in}} 
\author{Patricia Johann~~~~~~}
\author{Pierre Cagne}\vspace*{-0.1in}
\address{\em \textsf{~johannp@appstate.edu}~~~~~~~~~~\textsf{cagnep@appstate.edu}\vspace*{0.05in}\\  
Appalachian State University}

\vspace*{-0.1in}

\begin{abstract}
It is well-known that GADTs do not admit standard map functions of the
kind supported by ADTs and nested types. In addition, standard map
functions are insufficient to distribute their data-changing argument
functions over all of the structure present in elements of deep GADTs,
even just deep ADTs or nested types. This paper develops an algorithm
for detecting exactly which functions are mappable over data whose
types are (deep) GADTs. The algorithm takes as input a term $t$ whose
type is an instance of a deep GADT $\mathsf{D}$ and a function $f$ to
be mapped over $t$. It detects a minimal possible shape of $t$ as an
element of $\mathsf{D}$, and returns a minimal set of constraints $f$
must satisfy to be mappable over $t$. The crux of the algorithm is its
ability to separate $t$'s {\em essential structure} as an element of
$\mathsf{D}$ --- i.e., the part of $t$ that is essential for it to
have the shape of an element of $\mathsf{D}$ --- from its {\em
  incidental structure} as an element of $\mathsf{D}$ --- i.e., the
part of $t$ that is simply data in the positions of this shape. The algorithm ensures that the constraints on $f$ come
only from $t$'s essential structure.
This work is part of an ongoing effort to define initial algebra
semantics for GADTs that properly generalizes the usual semantics for
ADTs and nested types as least fixpoints of higher-order
endofunctors.\looseness=-1
\end{abstract}

\end{frontmatter}

\vspace*{-0.1in}

\section{Introduction}\label{sec:introduction}

%
%
%
%
%

\vspace*{-0.1in}

Initial algebra semantics~\cite{bdm97} is one of the cornerstones of
the modern theory of data types. It has long been known to deliver
practical programming tools --- such as pattern matching, induction
rules, and structured recursion operators --- as well as principled
reasoning techniques --- like relational parametricity~\cite{rey83}
--- for algebraic data types (ADTs). Initial algebra semantics has
also been developed for the syntactic generalization of ADTs known as
nested types~\cite{bm98}, and it has been shown to deliver analogous
tools and techniques for them as well~\cite{jg07}. Generalized
algebraic data types (GADTs)~\cite{pvww06,sp04,xcc04} generalize
nested types --- and thus further generalize ADTs --- syntactically:

\vspace*{-0.05in}

\begin{equation}\label{eq:hier}
\begin{tikzcd}[column sep = huge]
\text{\fbox{$\mathsf{ADTs}$}}\;
\ar[r,hookrightarrow,"\text{syntactically}","\text{generalized by}"']
& \;\text{\fbox{$\mathsf{nested~types}$}}\;
\ar[r,hookrightarrow,"\text{syntactically}","\text{generalized by}"'] 
& \;\text{\fbox{$\mathsf{GADTs}$}}
\end{tikzcd}
\end{equation}

\noindent
Given their ubiquity in modern functional programming, an important
open question is whether or not an initial algebra semantics exists
for GADTs.\looseness=-1

The starting point for initial algebra semantics is to interpret types
as objects in a suitably structured category $\mathcal C$, and to
interpret open type expressions as endofunctors on this category. An
ADT is interpreted as the least fixpoint of the endofunctor on
$\mathcal C$ interpreting its underlying type expression. For example,
the type expression underlying the standard data
type\footnote{Although our results apply to GADTs in any programming
  language, we will use Agda syntax for all code in this paper unless
  otherwise specified. But whereas Agda allows type parameters in the
  types of GADT data constructors to be implicit, we will always write
  all type parameters explicitly.  We use {\sf sans serif} font for
  code snippets and {\em italic} font for mathematics.}

\vspace*{-0.1in}

\begin{equation}\label{eq:lists}
\begin{array}{l}
\mathsf{data\; List \;:\;Set \to Set\;where}\\
\hspace*{0.2in}\mathsf{nil\;\;\;\;:\; \forall\,A \to List\;A}\\
\hspace*{0.2in}\mathsf{cons\;:\;\forall \,A \to A \rightarrow List\;A
  \rightarrow List\;A} 
\end{array}
\end{equation}

\vspace*{-0.02in}

\noindent
of lists of data of type $\mathsf{A}$ is $\mathsf{L_A \, X \,=\,1 + A
  \times X}$.  This is essentially the unfolding of the definition of
a type $\mathsf{X}$ parameterized on $\mathsf{A}$ recognizing that an
element of $\mathsf{X}$ can be constructed either from no data using
the data constructor $\mathsf{nil}$, or from one datum of type
$\mathsf{A}$ and one already-constructed datum of type $\mathsf{X}$
using the data constructor $\mathsf{cons}$. Replacing $\mathsf{X}$ by
$\mathsf{List\,A}$ in~\eqref{eq:lists} gives a recursive equation
defining this type, so if $A$ interprets $\mathsf{A}$ then the least
fixpoint of the endofunctor $L_A X = 1 + A \times X$ on $\mathcal C$
interpreting $\mathsf{L_A}$ interprets
$\mathsf{List\,A}$.\looseness=-1

\pagebreak

Nested types generalize ADTs by allowing their constructors to take as
arguments data whose types involve instances of the nested type other
than the one being defined. The return type of each of its data
constructors must still be precisely the instance being defined,
though. This is illustrated by the following standard definitions of
the nested types $\mathsf{PTree}$ of perfect trees and $\mathsf{Bush}$
of bushes:\looseness=-1

\vspace*{-0.25in}

\begin{equation*}\label{eq:ptrees}
\begin{array}{lll}
  \mathsf{data\; PTree \;:\;Set \to Set\;where} & \hspace*{0.3in} &
\mathsf{data\; Bush \;:\;Set \to Set\;where}\\
\hspace*{0.2in}\mathsf{pleaf\;\;\,:\; \forall\,A \to A \to PTree\;A} &
& \hspace*{0.2in}\mathsf{bnil\;\;\;\;:\; \forall\,A \to Bush\;A}\\
\hspace*{0.2in}\mathsf{pnode\;:\;\forall \,A \to PTree\;(A \times A)
  \rightarrow PTree\;A} & & \hspace*{0.2in}\mathsf{bcons\;:\;\forall
  \,A \to A \rightarrow  Bush\;(Bush\;A) \rightarrow Bush\;A}\\  
\end{array}
\end{equation*}

\vspace*{-0.13in}

\noindent
A nested type $\mathsf{N}$ with at least one data constructor at least
one of whose argument types involves an instance of $\mathsf{N}$ that
itself involves an instance of $\mathsf{N}$ is called a {\em truly
  nested type}. The type of the data constructor $\mathsf{bcons}$ thus
witnesses that $\mathsf{Bush}$ is a truly nested type.  Because the
recursive calls to a nested type's type constructor can be at
instances of the type other than the one being defined, a nested type
thus defines an entire family of types that must be constructed
simultaneously. That is, a nested type defines an {\em inductive
  family of types}. By contrast, an ADT is usually understood as a
family of inductive types, one for each choice of its type
arguments. This is because every recursive call to an ADT's type
constructor must be at the same instance as the one being
defined.\looseness=-1

Like ADTs, (truly) nested types can still be interpreted as least
fixpoints of endofunctors. But because the recursive calls in a nested
type's definition are not necessarily at the instance being defined,
the endofunctor interpreting its underlying type expression must
necessarily be a {\em higher-order} endofunctor on $\mathcal C$. For
example, the endofunctor interpreting the type expression underlying
$\mathsf{PTree}$ is $P\,F\,X = X + F\,(X \times X)$ and the
endofunctor interpreting the type expression underlying
$\mathsf{Bush}$ is $B\,F\,X = 1 + F\,(F\,X)$.  The fact that fixpoints
of higher-order endofunctors are themselves necessarily functors thus
entails that nested types are interpreted as endofunctors on, rather
than elements of, $\mathcal{C}$. This ensures that the fixpoint
interpretation of a nested type has a functorial action and, moreover,
that the map function for a nested type --- such as is required to
establish the nested type as an instance of Haskell's
$\mathsf{Functor}$ class\footnote{We write $\mathsf{map_{D}}$ for the
syntactic function $\mathsf{fmap :: (A \to B) \to (D\,A \to D\,B)}$
witnessing that the type constructor $\mathsf{D}$ is an instance of
Haskell's $\mathsf{Functor}$ class. Such functions are expected to
satisfy syntactic reflections of the functor laws --- i.e.,
preservation of identity functions and composition of functions ---
even though there is no compiler mechanism to enforce this.}  --- can
be obtained as its syntactic reflection. For example,
$\mathsf{map_{PTree}}$ is the syntactic reflection of the functorial
action of the fixpoint of $P$, and $\mathsf{map_{Bush}}$ is the
syntactic reflection of the functorial action of the fixpoint of
$B$. Because nested types, including ADTs and truly nested types, are
defined polymorphically, we can think of each element of such a type
$\mathsf{N}$ as a ``container'' for data arranged at various {\em
  positions} in the underlying {\em shape} determined by the data
constructors of $\mathsf{N}$ used to build it. Given a function
$\mathsf{f : A \to B}$, the function $\mathsf{map_N\,f}$ is then the
expected shape-preserving-but-possibly-data-changing function that
transforms an element of $\mathsf{N}$ with shape $S$ containing data
of type $\mathsf{A}$ into another element of $\mathsf{N}$ also of
shape $S$ but containing data of type $\mathsf{B}$ by applying
$\mathsf{f}$ to each of its elements.  The standard map functions for
ADTs can be obtained in the very same way --- i.e., by interpreting
them as fixpoints of (now trivially) higher-order endofunctors, rather
than of first-order endofunctors, on $\mathcal C$ and reflecting the
functorial actions of those fixpoints back into syntax. For example,
the usual map function $\mathsf{map_{List}}$ for lists is nothing more
than the syntactic reflection of the functorial action of the fixpoint
of the higher-order endofunctor $L'\,F\,X = 1 + X \times F\,X$
underlying $\mathsf{List}$.\looseness=-1

\label{page:seq}
Since GADTs syntactically subsume nested types, they would also
require higher-order endofunctors for their interpretation. We might
therefore expect GADTs to have {\em functorial initial algebra
  semantics}, and thus to support
shape-preserving-but-possibly-data-changing map functions, just like
nested types do. But because the shape of an element of a {\em proper}
GADT --- i.e., a GADT that is not a nested type (and thus is not an
ADT) --- is not independent of the data it contains, and is, in fact,
{\em determined by} this data, not all GADTs do. For example, the
GADT\looseness=-1

\vspace*{-0.2in}

\begin{equation*}\label{eq:seq}
\begin{array}{l}
\mathsf{data\; Seq \;:\;Set \to Set\;where}\\
\hspace*{0.3in}\mathsf{const\;:\; \forall A \to A \to Seq\,A}\\
\hspace*{0.3in}\mathsf{pair\;\;\;:\;\forall A\,B \to Seq\,A \to Seq\,B
  \rightarrow Seq\,(A \times B)}
\end{array}
\end{equation*}
of sequences does not support a standard
structure-preserving-but-possibly-data-changing map function like ADTs
and nested types do. If it did, then the clause of
$\mathsf{map_{Seq}}$ for an element of $\mathsf{Seq}$ of the form
$\mathsf{pair\,x\,y}$ for $\mathsf{x : A}$ and $\mathsf{y : B}$ would
be such that if $\mathsf{f : (A \times B) \to C}$ then
$\mathsf{map_{Seq}\,f\,(pair\,x\,y) = pair\,u\,v : Seq\,C}$ for some
appropriately typed $\mathsf{u}$ and $\mathsf{v}$. But there is no way
to achieve this unless $\mathsf{C}$ is of the form $\mathsf{A' \times
  B'}$ for some $\mathsf{A'}$ and $\mathsf{B'}$, $\mathsf{u :
  Seq\,A'}$ and $\mathsf{v : Seq\,B'}$, and $\mathsf{f = f_1 \times
  f_2}$ for some $\mathsf{f_1 : A \to A'}$ and $\mathsf{f_2 : B \to
  B'}$. The non-uniformity in the type-indexing of proper GADTs ---
which is the very reason a GADT programmer is likely to use GADTs in
the first place --- thus turns out to be precisely what prevents them
from supporting standard map functions.

Despite this, GADTs are currently known to support two different
functorial initial algebra semantics, namely, the discrete semantics
of~\cite{jg08} and the functorial completion semantics
of~\cite{jp19}. The problem is that neither of these leads to a
satisfactory uniform theory of type-indexed data types. On the one
hand, the discrete semantics of~\cite{jg08} interprets GADTs as
fixpoints of higher-order endofunctors on the {\em discretization} of
the category $\mathcal C$ interpreting types, rather than on $\mathcal
C$ itself. In this semantics, the map function for every GADT is
necessarily trivial. Viewing nested types as particular GADTs thus
gives a functorial initial algebra semantics for them that does not
coincide with the expected one. In other words, the discrete
interpretation of~\cite{jg08} results in a semantic situation that
does not reflect the syntactic one depicted in~\eqref{eq:hier}, and is
thus inadequate. On the other hand, the functorial completion
semantics of~\cite{jp19} interprets GADTs as endofunctors on $\mathcal
C$ itself. Each GADT thus, like every nested type, has a non-trivial
map function. This is, however, achieved at the cost of adding new
``junk'' elements, unreachable in syntax but interpreting elements in
the ``map closure'' of its syntax, to the interpretation of every
proper GADT.
Functorial completion for $\mathsf{Seq}$, e.g., adds interpretations
of elements of the form $\mathsf{map\,f\,(pair\,x\,y)}$ even though
these may not be of the form $\mathsf{pair\,u\,v}$ for any terms
$\mathsf{u}$ and $\mathsf{v}$. Importantly, functorial completion adds
no junk to interpretations of nested types or ADTs, so unlike the
semantics of~\cite{jg08}, that of~\cite{jp19} does indeed properly
extend the usual functorial initial algebra semantics for them. But
since the interpretations of~\cite{jp19} are bigger than expected for
proper GADTs, this semantics, too, is unacceptable. Although they are
at the two extremes of the junk vs.~functoriality spectrum, both known
functorial initial algebra semantics for GADTs are fundamentally
unsatisfactory.\looseness=-1

In this paper we pursue a middle ground and ask: how much
functoriality can we salvage for GADTs while still ensuring that their
interpretations contain no junk? We already know that not every
function on a proper GADT's type arguments will be mappable over
it. But this paper answers this question more precisely by developing
an algorithm for detecting exactly which functions are. Our algorithm
takes as input a term $t$ whose type is (an instance of) a GADT
$\mathsf{G}$ and a function $f$ to be mapped over $t$. It then
detects the {\em minimal possible shape} of $t$ as an element of
$\mathsf{G}$, and returns a {\em minimal set of constraints} $f$ must
satisfy in order to be mappable over $t$. The crux of the algorithm is
its ability to separate $t$'s {\em essential structure} as an element
of $\mathsf{G}$ --- i.e., the part of $t$ that is essential for it to
have the shape of an element of $\mathsf{G}$ --- from its {\em
  incidental structure} as an element of $\mathsf{G}$ --- i.e., the
part of $t$ that is simply data in the positions of this shape.  The
algorithm then ensures that the constraints ensuring that $f$ is
mappable come only from $t$'s essential structure as an element of
$\mathsf{G}$.

The separation of a term into essential and incidental structure
relative to a given specification is far from trivial, however. In
particular, it is considerably more involved than simply inspecting
the return types of $\mathsf{G}$'s constructors. As for ADTs and other
nested types, a subterm built using one of $\mathsf{G}$'s data
constructors can be an input term to another one (or to itself again),
and this creates a kind of ``feedback loop'' in the well-typedness
computation for the overall term. Moreover, if $\mathsf{G}$ is a
proper GADT, then such a loop can force structure to be essential in
the overall term even though it would be incidental in the subterm if
the subterm were considered in isolation, and this can impose
constraints on the functions mappable over it.  This is illustrated in
Examples~\ref{ex:ex2} and~\ref{ex:ex3} below, both of which involve a
GADT $\mathsf{G}$ whose data constructor $\mathsf{pairing}$ can
construct a term suitable as input to $\mathsf{projpair}$.

Our algorithm is actually far more flexible than we have just
described. Rather than simply considering $t$ to be an element of the
top-level GADT in its type, it can instead take as a third argument a
specification, in the form of a perhaps deeper\footnote{An ADT/nested
type/GADT is {\em deep} if it is (possibly mutually inductively)
defined in terms of other ADTs/nested types/GADTs (including,
possibly, itself). For example, $\mathsf{List\,(List\,\mathbb{N})}$ is
a deep ADT, $\mathsf{Bush\,(List\,(PTree\,A))}$ is a deep nested type,
and $\mathsf{Seq\,(PTree\,A)}$, and $\mathsf{List\,(Seq\,A)}$ are deep
GADTs.} data type $\mathsf{D}$, one of whose instances it should be
considered an element of. The algorithm will still return a minimal
set of constraints $f$ must satisfy in order to be mappable over $t$,
but now these constraints are relative to the deep specification
$\mathsf{D}$ rather than to the ``shallow'' specification
$\mathsf{G}\,\beta$. The feedback loops in and between the data types
appearing in the specification $\mathsf{D}$ can, however,
significantly complicate the separation of essential and incidental
structure in terms. For example, if a term's specification is
$\mathsf{G}\,(\mathsf{G}\,\beta)$ then we will first need to compute
which functions are mappable over its relevant subterms relative to
$\mathsf{G}\,\beta$ before we can compute those mappable over the term
itself relative to $\mathsf{G}\,(\mathsf{G}\,\beta)$. Runs of our
algorithm on deep specifications are given in Examples~\ref{ex:ex5}
and~\ref{ex:ex5-again} below, as well as in our accompanying code~\cite{code}.

This paper is organized as follows. Motivating examples highlighting
the delicacies of the problem our algorithm solves are given in
Section~\ref{sec:overview}. Our algorithm is given in
Section~\ref{sec:algorithm}, and fully worked out sample runs of it
are given in Section~\ref{sec:examples}. Our conclusions, related
work, and some directions for future work are discussed in
Section~\ref{sec:conclusion}. Our Agda implementation of our algorithm
is available at~\cite{code}, along with a collection of examples on
which it has been run. This collection includes examples involving
deep specifications and mutually recursively defined GADTs, as well as
other examples that go beyond just the illustrative ones appearing in
this paper.

\section{The Problem and Its Solution: An Overview}\label{sec:overview}

In this section we use well-chosen example instances of the mapping
problem for GADTs and deep data structures both to highlight its
subtlety and to illustrate the key ideas underlying our algorithm that
solves it. For each example considering a function $f$ to be mapped
over a term $t$ relative to the essential structure specified by $D$
we explain, intuitively, how to obtain the decomposition of $t$ into
the essential and incidental structure specified by $D$ and what the
minimal constraints are that ensure that $f$ is mappable over $t$
relative to it. By design, we handle the examples only informally in
this section. The results obtained by running our algorithm on their
formal representations are given in
Section~\ref{sec:examples}.\looseness=-1

Our algorithm will treat all GADTs in the class $\mathcal{G}$, whose
elements have the following general form when written in Agda:

\vspace*{-0.1in}

\begin{equation}\label{eq:gadts}
\begin{array}{l}
  \mathsf{data\ G : Set}^k
    \mathsf{\to Set\ where}\\
\mathsf{\;\;\;\;\;\;\;\;c}_1\, \mathsf{:\, t}_1\\
\;\;\;\;\;\;\;\;\vdots\\
\mathsf{\;\;\;\;\;\;\;\;c}_m\, \mathsf{:\, t}_m\\
\end{array}
\end{equation}
\noindent
where $k$ and $m$ can be any natural numbers, including $0$. Writing
$\ol v$ for a tuple $(v_1,...,v_l)$ when its length $l$ is clear from
context, and identifying a singleton tuple with its only element, each
data constructor $\mathsf{c}_i$, $i \in \{1,...,m\}$, has type
$\mathsf{t}_i$ of the form

\vspace*{-0.1in}

\begin{equation}\label{eq:data-constr-types}
\mathsf{\forall \ol\alpha \to\,}
F_1^{\mathsf{c}_i} \mathsf{\ol \alpha \to ... \to\,}
F_n^{\mathsf{c}_i} \mathsf{\ol \alpha \to G\,}
(K_1^{\mathsf{c}_i} \mathsf{\ol \alpha, ... ,}
K_k^{\mathsf{c}_i} \mathsf{\ol\alpha)} 
\end{equation}
\noindent
Here, for each $j \in \{1,...,n\}$, $F_j^{\,\mathsf{c}_i}\ol\alpha$ is
either a closed type, or is $\alpha_d$ for some $d \in
\{1,...,|\ol\alpha|\}$, or is $\mathsf{D}_j^{\mathsf{c}_i}
\,(\ol{\phi_j^{\mathsf{c}_i}\ol\alpha})$ for some user-defined data
type constructor $\mathsf{D}_j^{\mathsf{c}_i}$ and tuple
$\ol{\phi_j^{\mathsf{c}_i}\ol\alpha}$ of type expressions at least one
of which is not closed. The types $F_j^{\,\mathsf{c}_i}\ol\alpha$ must
not involve any arrow types.  However, each
$\mathsf{D}_j^{\mathsf{c}_i}$ can be any GADT in $\mathcal{G}$,
including $\mathsf{G}$ itself, and each of the type expressions in
$\ol{\phi_j^{\mathsf{c}_i}\ol\alpha}$ can involve such GADTs as
well. On the other hand, for each $\ell \in \{1,...,k\}$,
$K_\ell^{\mathsf{c}_i} \ol\alpha$ is a type expression whose free
variables come from $\ol\alpha$, and that involves neither $G$ itself
nor any proper GADTs.\footnote{Formally, a GADT is a {\em proper} GADT
if it has at least one {\em restricted data constructor}, i.e., at
least one data constructor $\mathsf{c}_i$ with type as
in~\eqref{eq:data-constr-types} for which $K_\ell^{\mathsf{c}_i} \ol
\alpha \not = \ol \alpha$ for at least one $\ell \in \{1,...,k\}$.}
When $|\ol \alpha| = 0$ we suppress the initial quantification over
types in~\eqref{eq:data-constr-types}. All of the GADTs appearing in
this paper are in the class $\mathcal{G}$. All GADTs we are aware of
from the literature whose constructors' argument types do not involve
arrow types are also in $\mathcal{G}$. Our algorithm is easily
extended to GADTs without this restriction provided all arrow types
involved are strictly positive.\looseness=-1

Our first example picks up the discussion for $\mathsf{Seq}$ on
page~\pageref{page:seq}. Because $\mathsf{pair}$ is the only
restricted data constructor for $\mathsf{Seq}$, so that the feedback
dependencies for $\mathsf{Seq}$ are simple, it is entirely
straightforward.

\begin{example}\label{ex:ex1}
The functions $f$ mappable over

\vspace*{-0.05in}

\begin{equation}
t = \mathsf{pair}\,(\mathsf{ pair}\,(\mathsf{{const}\,tt})\,
(\mathsf{{ const}\,2}))\;(\mathsf{{const}\,5}) :
\mathsf{Seq}\,(\,(\mathsf{Bool} \times \mathsf{Int}) \times
\mathsf{Int})
\end{equation}

\vspace*{0.1in}

\noindent
relative to the specification $\mathsf{Seq}\,\beta$ are exactly those
of the form $f = (f_1 \times f_2) \times f_3$ for some $f_1 :
\mathsf{Bool} \to X_1$, $f_2 : \mathsf{Int} \to X_2$, and $f_1 :
\mathsf{Int} \to X_3$, and some types $X_1$, $X_2$, and
$X_3$. Intuitively, this follows from two analyses similar to that on
page~\pageref{page:seq}, one for each occurrence of $\mathsf{pair}$ in
$t$. Writing the part of a term comprising its essential structure
relative to the given specification in {\color{blue} blue} and the
parts of the term comprising its incidental structure in black, our
algorithm also deduces the following essential structure for $t$:

\vspace*{-0.1in}

\[\mathsf{\color{blue} pair}\,(\mathsf{\color{blue}
  pair}\,(\mathsf{{\color{blue} const}\,tt})\, (\mathsf{{\color{blue}
    const}\,2}))\;(\mathsf{{\color{blue} const}\,5}) :
\mathsf{Seq}\,(\,(\mathsf{Bool} \times \mathsf{Int}) \times
\mathsf{Int})\]

\vspace*{-0.1in}

\end{example}

The next two examples are more involved because $\mathsf{G}$ has
purposely been crafted so that its data constructor $\mathsf{pairing}$
can construct a term suitable as input to $\mathsf{projpair}$.

\begin{example}\label{ex:ex2}
Consider the GADT 

\vspace*{-0.1in}

\begin{equation*}\label{eq:crazy}
\begin{array}{l}
\mathsf{data\; G\;:\;Set \to Set\;where}\\
\hspace*{0.3in}\mathsf{const\;\;\;\;:\; G\,\mathbb{N}}\\
\hspace*{0.3in}\mathsf{flat\;\;\;\;\;\;\;:\;\forall\,A \to
  List\,(G\,A) \to G\,(List\,A)}\\ 
\hspace*{0.3in}\mathsf{inj\;\;\;\;\;\;\;\;\,:\; \forall\,A \to A \to
  G\,A}\\ 
\hspace*{0.3in}\mathsf{pairing\;\;:\;\forall\,A\,B \to G\,A \to G\,B
  \rightarrow G\,(A \times B)}\\
\hspace*{0.3in}\mathsf{projpair\;:\;\forall\,A\,B \to G\,(G\,A \times
  G\,(B \times B)) \to G\,(A \times B)}
\end{array}
\end{equation*}
The functions mappable over

\vspace*{-0.1in}

\begin{equation*}\label{eq:u-inf}
t = {\mathsf{projpair}} \;(\;{\mathsf{inj}\;} \;
(\mathsf{{inj}\;(cons\,2\,nil)},\; {\mathsf{pairing}}\; ({\mathsf{
    inj}\,2})\;{\mathsf{const}}) \;) :
\mathsf{G}\,(\mathsf{List}\,\mathbb{N} \, \times \, \mathbb{N})
\end{equation*}
relative to the specification $\mathsf{G}\,\beta$ are exactly those of
the form $f = f_1 \times \id_\mathbb{N}$ for some type $X$ and
function $f_1 : \mathsf{List}\,\mathbb{N} \to X$. This makes sense
intuitively: The call to $\mathsf{projpair}$ requires that a mappable
function $f$ must at top level be a product $f_1 \times f_2$ for some
$f_1$ and $f_2$, and the outermost call to $\mathsf{inj}$ imposes no
constraints on $f_1 \times f_2$. In addition, the call to
$\mathsf{inj}$ in the first component of the pair argument to the
outermost call to $\mathsf{inj}$ imposes no constraints on $f_1$, and
neither does the call to $\mathsf{cons}$ or its arguments. On the
other hand, the call to $\mathsf{pairing}$ in the second component of
the pair argument to the second call to $\mathsf{inj}$ must produce a
term of type $\mathsf{G\,(\mathbb{N} \times \mathbb{N})}$, so the
argument $\mathsf{2}$ to the rightmost call to $\mathsf{inj}$ and the
call to $\mathsf{const}$ require that $f_2 = \id_{\mathbb{N}}$. Our
algorithm also deduces the following essential structure for $t$:

\vspace*{-0.1in}

\begin{equation}\label{ex:decomp}
{\mathsf{\color{blue} projpair}} \;(\;{\mathsf{\color{blue}
    inj}\;} \; (\mathsf{{\color{blue} inj}\;(cons\,2\,nil)},\;
  {\mathsf{\color{blue} pairing}}\; ({\mathsf{\color{blue}
      inj}\,2})\;{\mathsf{\color{blue} const}}) \;) :
  \mathsf{G}\,(\mathsf{List}\,\mathbb{N} \, \times \, \mathbb{N})
\end{equation}  
Note that, although the argument to $\mathsf{projpair}$ decomposes
into essential structure and incidental structure as
$\mathsf{\color{blue}inj}\,(\mathsf{{inj}\;(cons\,2\,nil)},\;
{\mathsf{pairing}}\; ({\mathsf{ inj}\,2})\;{\mathsf{const}})$ when
considered as a standalone term relative to the specification
$\mathsf{G}\,\beta$, the feedback loop between $\mathsf{pairing}$ and
$\mathsf{projpair}$ ensures that $t$ has the decomposition in
\eqref{ex:decomp} relative to $\mathsf{G}\,\beta$ when this argument
is considered in the context of $\mathsf{projpair}$. Similar comments
apply throughout this paper.
\end{example}

\begin{example}\label{ex:ex3}
The functions $f$ mappable over

\vspace*{-0.1in}

\begin{equation*}\label{eq:t-inf}
t = {\mathsf{projpair}}\;(\;{\mathsf{inj}\;} \;
({\mathsf{flat}}\;({\mathsf{ cons}}\;{\mathsf{const}}\;{\mathsf{
    nil}}),\; {\mathsf{pairing}}\; ({\mathsf{
    inj}\,2})\;{\mathsf{const}}) \;) :
\mathsf{G}\,(\mathsf{List}\,\mathbb{N} \, \times \, \mathbb{N})
\end{equation*}
relative to the specification $\mathsf{G}\,\beta$ for $\mathsf{G}$ as
in Example~\ref{ex:ex2} are exactly those of the form $f =
map_{\mathsf{List}}\,\id_{\mathbb{N}} \times \id_\mathbb{N}$.  This
makes sense intuitively: The call to $\mathsf{projpair}$ requires that
a mappable function $f$ must at top level be a product $f_1 \times
f_2$ for some $f_1$ and $f_2$, and the outermost call to
$\mathsf{inj}$ imposes no constraints on $f_1 \times f_2$. In
addition, the call to $\mathsf{flat}$ in the first component of the
pair argument to $\mathsf{inj}$ requires that $f_1 =
map_{\mathsf{List}}\,f_3$ for some $f_3$, and the call to
$\mathsf{cons}$ in $\mathsf{flat}$'s argument imposes no constraints
on $f_3$, but the call to $\mathsf{const}$ as $\mathsf{cons}$'s first
argument requires that $f_3 = \id_{\mathbb{N}}$. On the other hand, by
the same analysis as in Example~\ref{ex:ex2}, the call to
$\mathsf{pairing}$ in the second component of the pair argument to
$\mathsf{inj}$ requires that $f_2 = \id_{\mathbb{N}}$. Our algorithm
also deduces the following essential structure for $t$:

\vspace*{-0.1in}

\[{\mathsf{\color{blue}
    projpair}}\;(\;{\mathsf{\color{blue} inj}\;} \;
({\mathsf{\color{blue} flat}}\;({\mathsf{\color{blue}
    cons}}\;{\mathsf{\color{blue} const}}\;{\mathsf{\color{blue}
    nil}}),\; {\mathsf{\color{blue} pairing}}\; ({\mathsf{\color{blue}
    inj}\,2})\;{\mathsf{\color{blue} const}}) \;) :
\mathsf{G}\,(\mathsf{List}\,\mathbb{N} \, \times \, \mathbb{N})\]
\end{example}

The feedback loop between constructors in the GADT $\mathsf{G}$ in the
previous two examples highlights the importance of the specification
relative to which a term is considered. But this can already be seen
for ADTs, which feature no such loops. This is illustrated in
Examples~\ref{ex:ex4} and~\ref{ex:ex5} below.

\begin{example}\label{ex:ex4}
The functions $f$ mappable over

\vspace*{-0.1in}

\begin{equation*}
t = \mathsf{cons}\,(\mathsf{cons\,1\,(cons\,2\,nil))\,
  (cons\,(cons\,3\,nil)\,nil) : List\,(List\,\mathbb{N})}
\end{equation*}
relative to the specification $\mathsf{List\,\beta}$ are exactly those
of the form $f : \mathsf{List}\,\mathbb{N} \to X$ for some type
$X$. This makes sense intuitively since any function from the element
type of a list to another type is mappable over that list. The
function need not satisfy any particular structural constraints. Our
algorithm also deduces the following essential structure for $t$:
\[\mathsf{{\color{blue}cons}}\,(\mathsf{cons\,1\,(cons\,2\,nil))\,
  ({\color{blue}cons}\,(cons\,3\,nil)\,{\color{blue}nil})}\]
\end{example}

\begin{example}\label{ex:ex5}
  The functions $f$ mappable over

  \vspace*{-0.1in}
  
\begin{equation*}
t = \mathsf{cons\,(cons\,1\,(cons\,2\, nil))\,(cons\,(cons\,3\,
  nil)\,nil) : List\,(List\,\mathbb{N})}
\end{equation*}
relative to the specification $\mathsf{List\, (List\,\beta)}$ are
exactly those of the form $f = \mathsf{map_{List}}\,f'$ for some type
$X'$ and function $f' : \mathbb{N} \to X'$. This makes sense
intuitively: The fact that any function from the element type of a
list to another type is mappable over that list requires that $f :
\mathsf{List}\,\mathbb{N} \to X$ for some type $X$ as in
Example~\ref{ex:ex4}. But if the internal list structure of $t$ is
also to be preserved when $f$ is mapped over it, as indicated by the
essential structure $\mathsf{List}\,(\mathsf{List}\,\beta)$, then $X$ must
itself be of the form $\mathsf{List}\,X'$ for some type $X'$. This, in
turn, entails that $f = map_{\mathsf{List}}f'$ for some $f' :
\mathbb{N} \to X'$. Our algorithm also deduces the following essential
structure for $t$:\looseness=-1

\vspace*{-0.1in}

\[\mathsf{{\color{blue}cons}\,({\color{blue}cons}\,1\,({\color{blue}cons}\,2\,
  {\color{blue}nil}))\,({\color{blue}cons}\,({\color{blue}cons}\,3\,
  {\color{blue}nil})\,{\color{blue}nil}) : List\,(List\,\mathbb{N})}\]
\end{example}

The specification $\mathsf{List}\,(\mathsf{List}\,\beta)$ determining
the essential structure in Example~\ref{ex:ex5} is deep {\em by
  instantiation}, rather than {\em by definition}. That is, inner
occurrence of $\mathsf{List}$ in this specification is not forced by
the definition of the data type $\mathsf{List}$ that specifies its
top-level structure. The quintessential example of a data type that is
deep by definition is the ADT\looseness=-1

\vspace*{-0.15in}

\begin{equation*}\label{eq:rose}
\begin{array}{l}
\mathsf{data\; Rose \;:\;Set \to Set\;where}\\
\hspace*{0.2in}\mathsf{rnil\;\;\;\;:\; \forall\,A \to Rose\;A}\\
\hspace*{0.2in}\mathsf{rnode\;:\;\forall \,A \to A \to List\,(Rose\,A)
  \to Rose\;A} 
\end{array}
\end{equation*}
of rose trees, whose data constructor $\mathsf{rnode}$ takes as input
an element of $\mathsf{Rose}$ at an instance of another ADT.
Reasoning analogous to that in the examples above suggests that no
structural constraints should be required to map appropriately typed
functions over terms whose specifications are given by nested types
that are deep by definition.  We will see in Example~\ref{ex:list}
that, although the runs of our algorithm are not trivial on such input
terms, this is indeed the case.

With more tedious algorithmic bookkeeping, results similar to those of
the above examples can be obtained for data types --- e.g.,
$\mathsf{Bush\,(List\,(PTree\,A))}$, $\mathsf{Seq\,(PTree\,A)}$, and
$\mathsf{List\,(Seq\,A)}$ --- that are deep by instantiation~\cite{code}.

\section{The Algorithm}\label{sec:algorithm}

In this section we give our algorithm for detecting mappable
functions. The algorithm $\mathsf{adm}$ takes as input a data
structure $t$, a tuple of functions to be mapped over $t$, and a
specification --- i.e., a (deep) data type --- $\Phi$. It detects the
minimal possible shape of $t$ relative to $\Phi$ and returns a minimal
set $C$ of constraints $\ol f$ must satisfy in order to be mappable
over $t$ viewed as an element of an instance of $\Phi$.  A call

\vspace*{-0.1in}

\[\mathsf{adm}\;\:t\;\:\ol f\;\;\Phi\]
is to be made only if there exists a tuple $(\Sigma_1
\ol\beta,...,\Sigma_k\ol\beta)$ of type expressions such that
\begin{itemize}
\item $\Phi = \mathsf{G}\,(\Sigma_1 \ol\beta,...,\Sigma_k\ol\beta)$
  for some data type constructor $\mathsf{G} \in \mathcal{G} \cup
  \{{\times},{+}\}$ and some type expressions $\Sigma_\ell \ol\beta$,
  for $\ell \in\{1,...,k\}$
\end{itemize}
and 
\begin{itemize}
\item if $\Phi = \times (\Sigma_1 \ol \beta, \Sigma_2 \ol \beta)$,
  then $t = (t_1, t_2)$, and $k = 2$, $\ol f = (f_1, f_2)$
\item if $\Phi = + (\Sigma_1 \ol \beta, \Sigma_2 \ol \beta)$ and $t =
  \mathsf{inl}\,t_1$, then $k = 2$, $\ol f = (f_1, f_2)$
\item if $\Phi = + (\Sigma_1 \ol \beta, \Sigma_2 \ol \beta)$ and $t =
  \mathsf{inr}\,t_2$, then $k = 2$, $\ol f = (f_1, f_2)$
\item if $\Phi = \mathsf{G}\,(\Sigma_1 \ol\beta,...,\Sigma_k\ol\beta)$
for some $\mathsf{G} \in \mathcal{G}$ then
\begin{enumerate}[label=\arabic*)]
\item $t = \mathsf{c}\,t_1 ... t_n$ for some appropriately typed terms
$t_1,...,t_n$ and some data constructor $\mathsf{c}$ for $\mathsf{G}$
with type of the form in~\eqref{eq:data-constr-types},
\item \label{page:5b} $t : \mathsf{G}\,(K_1^\mathsf{c} \ol w, ... ,
  K_k^\mathsf{c} \ol w)$ for some tuple $\ol w =
  (w_1,...,w_{|\ol\alpha|})$ of type expressions, and
  $\mathsf{G}\,(K_1^\mathsf{c} \ol w, ... , K_k^\mathsf{c} \ol w)$ is
  exactly $\mathsf{G}\,(\Sigma_1 \ol s,...,\Sigma_k \ol s)$ for some
  tuple $\ol s = (s_1,...,s_{|\ol\beta|})$ of types, and\label{invariant2}
\item for each $\ell \in \{1,...,k\}$, $f_\ell$ has domain
  $K_\ell^\mathsf{c} \ol w$ 
\end{enumerate}
\end{itemize}
These invariants are clearly preserved for each recursive call to
$\mathsf{adm}$. 

As an optimization, the free variables in the type expressions
$\Sigma_\ell\ol\beta$ for $\ell \in \{1,...,k\}$ can be taken merely
to be {\em among} the variables in $\ol \beta$, since the calls
$\mathsf{adm}\;t\;\;\ol f\;\;\mathsf{G}\,(\Sigma_1\ol\beta,...,
\Sigma_k\ol\beta)$ and $\mathsf{adm}\;t\;\;\ol
f\;\;\mathsf{G}\,(\Sigma_1\ol{\beta^+},..., \Sigma_k\ol{\beta^+})$
return the same set $C$ (up to renaming) whenever $\ol{\beta}$ is a
subtuple of the tuple $\ol{\beta^+}$. We therefore always take $\ol
\beta$ to have minimal length below.

The algorithm is given as follows by enumerating each of its legal
calls.  Each call begins by initializing a set $C$ of constraints to
$\emptyset$.
\begin{itemize}
\item[A.]
 $\mathsf{adm}\;(t_1,t_2)\;\;(f_1,f_2)\;\;{\times} (\Sigma_1\ol\beta,
  \Sigma_2\ol\beta)$
\begin{enumerate}
\item Introduce a tuple $\ol g = g_1,...,g_{|\ol\beta|}$ of fresh
  function variables, and add the
  constraints $\lop \Sigma_1 \ol g, f_1 \rop$ and $\lop \Sigma_2 \ol
  g, f_2 \rop$ to $C$.
\item For $j \in \{1,2\}$,
  if $\Sigma_j\ol\beta = \beta_i$ for some $i$ then do nothing and
  go to the next $j$ if there is one. Otherwise, $\Sigma_j\ol\beta =
  \mathsf{D}\, (\zeta_1\ol\beta,...,\zeta_r\ol\beta)$, where
  $\mathsf{D}$ is a data type constructor in $\mathcal{G} \cup
  \{{\times},{+}\}$ of arity $r$, so make the recursive call
  $\mathsf{adm}\;t_j\;\;(\zeta_1\ol g,...,\zeta_r\ol
  g)\;\;\mathsf{D}\, (\zeta_1\ol\beta,...,\zeta_r\ol\beta)$ and add
  the resulting constraints to $C$.
\item Return $C$.
\end{enumerate}
\item[B.] $\mathsf{adm}\; (\mathsf{inl}\, t)\;\; (f_1 , f_2)\;\;
    {+} (\Sigma_1\ol\beta, \Sigma_2\ol\beta)$
\begin{enumerate}
\item Introduce a tuple $\ol g = (g_1,...,g_{|\ol\beta|})$ of fresh
function variables, and add the constraints $\lop \Sigma_1 \ol g,
    f_1 \rop$ and $\lop \Sigma_2 \ol g, f_2 \rop$ to $C$.
\item If $\Sigma_1\ol\beta = \beta_i$ for some $i$ then do nothing.
  Otherwise, $\Sigma_1\ol\beta = \mathsf{D}\,
  (\zeta_1\ol\beta,...,\zeta_r\ol\beta)$, where $\mathsf{D}$ is a data
  type constructor in $\mathcal{G} \cup \{{\times},{+}\}$ of arity
  $r$, so make the recursive call $\mathsf{adm}\;t\;\;(\zeta_1\ol
  g,...,\zeta_r\ol g)\;\;\mathsf{D}\,
  (\zeta_1\ol\beta,...,\zeta_r\ol\beta)$ and add the resulting
  constraints to $C$.
\item Return $C$.
\end{enumerate}
\item[C.] $\mathsf{adm}\; (\mathsf{inr}\, t)\;\; (f_1 , f_2)\;\;
    {+} (\Sigma_1\ol\beta, \Sigma_2\ol\beta)$
\begin{enumerate}
\item Introduce a tuple $\ol g = (g_1,...,g_{|\ol\beta|})$ of fresh
function variables, and add the constraints $\lop \Sigma_1 \ol g,
    f_1 \rop$ and $\lop \Sigma_2 \ol g, f_2 \rop$ to $C$.
\item If $\Sigma_2\ol\beta = \beta_i$ for some $i$ then do nothing.
  Otherwise, $\Sigma_2\ol\beta = \mathsf{D}\,
  (\zeta_1\ol\beta,...,\zeta_r\ol\beta)$, where $\mathsf{D}$ is a data
  type constructor in $\mathcal{G} \cup \{{\times},{+}\}$ of arity
  $r$, so make the recursive call $\mathsf{adm}\;t\;\;(\zeta_1\ol
  g,...,\zeta_r\ol g)\;\;\mathsf{D}\,
  (\zeta_1\ol\beta,...,\zeta_r\ol\beta)$ and add the resulting
  constraints to $C$.
\item Return $C$.
\end{enumerate}
\item[D.] $\mathsf{adm}\; (\mathsf{c}\, t_1,...,t_n)\;\; (f_1,...,f_k)\;\;
  \mathsf{G}\, (\Sigma_1 \ol\beta,...,\Sigma_k \ol\beta)$
\begin{enumerate}
\item Introduce a tuple $\ol g = (g_1,...,g_{|\ol\beta|})$ of fresh
  function variables and add the constraints $\lop \Sigma_\ell \ol g,
  f_\ell \rop$ to $C$ for each $\ell \in \{1,...,k\}$.
\item If $\mathsf{c}\, t_1,...,t_n : \mathsf{G}\,(K_1^\mathsf{c} \ol
  w,..., K_k^\mathsf{c} \ol w)$ for some tuple $\ol w =
  (w_1,...,w_{|\ol\alpha|})$ of types, let $\ol\gamma =
  (\gamma_1,...,\gamma_{|\ol\alpha|})$ be a tuple of fresh type
  variables and solve the system of matching problems

\vspace*{-0.2in}

\begin{align*}
 &\Sigma_1\ol\beta \equiv K^\mathsf{c}_1\ol\gamma\\
 &\Sigma_2\ol\beta \equiv K^\mathsf{c}_2\ol\gamma\\
 &\vdots\\
 &\Sigma_k\ol\beta \equiv K^\mathsf{c}_k\ol\gamma
\end{align*}

\vspace*{-0.1in}

\noindent
to get a set of assignments, each of the form $\beta \equiv
\psi\ol\gamma$ or $\sigma \ol \beta \equiv \gamma$ for some type
expression $\psi$ or $\sigma$. This yields a (possibly empty) tuple of
assignments $\ol {\beta_i \equiv \psi_i\ol\gamma}$ for each
$i\in\{1,...,|\ol\beta|\}$, and a (possibly empty) tuple of
assignments $\ol {\sigma_{i'}\ol\beta \equiv \gamma_{i'}}$ for each
$i'\in\{1,\dots,\len{\ol\gamma}\}$. Write $\beta_i\equiv
\psi_{i,p}\ol\gamma$ for the $p^{th}$ component of the former and
$\sigma_{i',q}\ol\beta \equiv \gamma_{i'}$ for the $q^{th}$ component
of the latter. An assignment $\beta_i \equiv \gamma_{i'}$ can be seen
as having form $\beta_i \equiv \psi\ol\gamma_{i'}$ or form
$\sigma\ol\beta_i \equiv \gamma_{i'}$, but always choose the latter
representation.  (This is justified because $\mathsf{adm}$ would
return an equivalent set of assignments --- i.e., a set of assignments
yielding the same requirements on $\ol f$ --- were the former
chosen. The latter is chosen because it may decrease the number of
recursive calls to $\mathsf{adm}$.)
\item For each $i' \in\{1,\dots,\len{\ol\gamma}\}$, define
  $\tau_{i'}\ol\beta\ol\gamma$ to be either
  $\sigma_{i',1}\ol\beta$ if this exists, or $\gamma_{i'}$ otherwise.
\item Introduce a tuple $\ol h = (h_1,...,h_{\len{\ol\gamma}})$ of fresh
  function variables for $i' \in \{1,...,|\ol\gamma|\}$.
\item For each $i\in\{1,\dots,\len{\ol\beta}\}$ and each
    constraint $\beta_i \equiv \psi_{i,p}\ol\gamma$, add the
    constraint $\lop \psi_{i,p}\ol h , g_i \rop$ to $C$.
\item For each $i'\in\{1,\dots,\len{\ol\gamma}\}$ and each constraint
  $\sigma_{i',q}\ol\beta \equiv \gamma_{i'}$ with $q>1$, add the
  constraint $\lop \sigma_{i',q}\ol g , \sigma_{i',1}\ol g \rop$ to $C$.
\item For each $j\in \{1,\dots,n\}$, let $R_j = F^\mathsf{c}_j
  (\tau_1\ol\beta\ol\gamma,...,\tau_{\len{\ol\gamma}}\ol\beta\ol\gamma)$.
\begin{itemize}[label=--]
\item if $R_j$ is a closed type, then do nothing and go to the next
  $j$ if there is one.
\item if $R_j = \beta_i$ for some $i$ or $R_j = \gamma_{i'}$ for some
  $i'$, then do nothing and go to the next $j$ if there is one.
\item otherwise $R_j = \mathsf{D}\,
  (\zeta_{j,1}\ol\beta\ol\gamma,...,\zeta_{j,r}\ol\beta\ol\gamma)$, where
  $\mathsf{D}$ is a type constructor in $\mathcal{G}\; \cup\;
  \{{\times},{+}\}$ of arity $r$, so make the recursive call

\vspace*{-0.15in}
  
\[\mathsf{adm}\;t_j\;\;(\zeta_{j,1}\ol g \ol h, ... , \zeta_{j,r}\ol
g\ol h)\;\; R_j\]

\noindent
and add the resulting constraints to $C$.
    \end{itemize}
\item Return $C$.
\end{enumerate}
\end{itemize}

We note that the matching problems in Step~(ii) in the last bullet
point above do indeed lead to a set of assignments of the specified
form. Indeed, since invariant \ref{invariant2} on
page~\pageref{page:5b} ensures that $\mathsf{G}\,(K_1^\mathsf{c} \ol
w, ... , K_k^\mathsf{c} \ol w)$ is exactly $\mathsf{G}\,(\Sigma_1 \ol
s,...,\Sigma_k \ol s)$, each matching problem $\Sigma_\ell \ol\beta
\equiv K_\ell \ol\gamma$ whose left- or right-hand side is not already
just one of the $\beta$s or one of the $\gamma$s must necessarily have
left- and right-hand sides that are {\em top-unifiable}~\cite{dj92},
i.e., have identical symbols at every position that is a non-variable
position in both terms.  These symbols can be simultaneously peeled
away from the left- and right-hand sides to decompose each matching
problem into a unifiable set of assignments of one of the two forms
specified in Step~(ii). We emphasize that the set of assignments is
not itself unified in the course of running $\mathsf{adm}$.

It is only once $\mathsf{adm}$ is run that the set of constraints it
returns is to be solved. Each such constraint must be either of the
form $\lop \Sigma_\ell \ol g, f_\ell \rop$, of the form
$\lop \psi_{i,p}\ol h , g_i \rop$, or of the form
$\lop \sigma_{i',q}\ol g , \sigma_{i',1}\ol g \rop$. Each constraint
of the first form must have top-unifiable left- and right-hand
components by virtue of invariant \ref{invariant2} on
page~\pageref{page:5b}. It can therefore be decomposed in a manner
similar to that described in the preceding paragraph to arrive at a
unifiable set of constraints. Each constraint of the second form
simply assigns a replacement expression $\psi_{i,p}\ol h$ to each
newly introduced variable $g_i$. Each constraint of the third form
must again have top-unifiable left- and right-hand components. Once
again, invariant \ref{invariant2} on page~\pageref{page:5b} ensures that
these constraints are decomposable into a unifiable set of constraints
specifying replacement functions for the $g$s.\looseness=-1

Performing first-order unification on the entire system of constraints
resulting from the decompositions specified above, and choosing to
replace more recently introduced $g$s and $h$s with ones introduced
later whenever possible, yields a {\em solved system} comprising
exactly one binding for each of the $f$s in terms of those
later-occurring variables. These bindings actually determine the
collection of functions mappable over the input term to $\mathsf{adm}$
relative to the specification $\Phi$. It is not hard to see that our
algorithm delivers the expected results for ADTs and nested types
(when $\Phi$ is the type itself), namely, that all appropriately typed
functions are mappable over each elements of such types. (See
Theorem~\ref{thm:ext} below.)  For GADTs, however, there is no {\em
  existing} understanding of which functions should be mappable over
their terms. We therefore regard the solved system's bindings for the
$f$s as actually {\em defining} the class of functions mappable over a
given term relative to a specification $\Phi$.

\begin{theorem}\label{thm:ext}
Let $\mathsf{N}$ be a nested type of arity $k$ in $\mathcal{G}$, let
$\ol w = (w_1,\ldots, w_k)$ comprise instances of nested types in
$\mathcal{G}$, let $t : \mathsf{N}\,\ol w$ where $\mathsf{N}\,\ol w$
contains $n$ free type variables, let $\ol\beta =
(\beta_1,\ldots,\beta_n)$, and let $\mathsf{N}\,(\Sigma_1 \ol\beta,
\ldots, \Sigma_k \ol\beta)$ be in $\mathcal{G}$. The solved system
resulting from the call $~\mathsf{adm}\;\;t\;\;(\Sigma_1 \ol
f,\ldots,\Sigma_k \ol f)\;\;\mathsf{N}\,(\Sigma_1 \ol\beta, \ldots,
\Sigma_k \ol\beta)$ for $\ol f = (f_1,\ldots,f_n)$ has the form
$\bigcup_{i=1}^n\{\lop g_{i,1} ,f_i \rop, \lop g_{i,2}, g_{i,1} \rop,
\ldots, \lop g_{i,r_i-1}, g_{i,r_i} \rop \}$, where each $r_i \in
\mathbb{N}$ and the $g_{i,j}$ are pairwise distinct function
variables. It thus imposes no constraints on the functions mappable
over terms of ADTs and nested types.\looseness=-1
\end{theorem}
\begin{proof}
The proof is by cases on the form of the given call to
$\mathsf{adm}$. The constraints added to $\mathcal C$ if this call is
of the form A, B, or C are all of the form $\lop \Sigma_j \ol g,
\Sigma_j\ol f \rop$ for $j = 1, 2$, and the recursive calls made are
all of the form $\mathsf{adm}\;t'\;\;(\zeta_1\ol g,..., \zeta_r\ol
g)\;\;\mathsf{D}\, (\zeta_1\ol\beta,...,\zeta_r\ol\beta)$ for some
$t'$, some $(\zeta_1,...,\zeta_r)$, and some nested type
$\mathsf{D}$. Now suppose the given call is of the form D. Then Step
(i) adds the constraints $\lop \Sigma_i\ol g, \Sigma_i \ol f \rop$ for
$i = 1,\dots,k$ to $\mathcal C$. In Step (ii), $|\ol\alpha| = k$, and
$K^{\mathsf{c}}_i \ol w = w_i$ for $i = 1,\ldots,k$ for every data
constructor $\mathsf{c}$ for every nested type, so that the matching
problems to be solved are $\Sigma_i\ol\beta \equiv \gamma_i$ for $i =
1,\ldots,k$. In Step (iii) we therefore have $\tau_i\ol\beta\ol\gamma
= \Sigma_i \ol\beta$ for $i = 1,\ldots,k$. No constraints involving
the variables $\ol h$ introduced in Step (iv) are added to $\mathcal
C$ in Step (v), and no constraints are added to $\mathcal C$ in Step
(vi) since the $\gamma$s are all fresh and therefore pairwise
distinct. For each $R_j$ that is of the form $\mathsf{D}\,
(\zeta_{j,1}\ol\beta\ol\gamma, \ldots ,\zeta_{j,r}\ol\beta\ol\gamma)$,
where $\mathsf{D}$ is a nested type, the recursive call added to
$\mathcal C$ in Step (vii) is of the form
$\mathsf{adm}\;\;t_j\;\;(\zeta_{j,1}\ol g \ol h, \ldots
,\zeta_{j,r}\ol g \ol h)\;\;\mathsf{D}\,(\zeta_{j,1}\ol\beta\ol\gamma,
\ldots ,\zeta_{j,r}\ol\beta\ol\gamma)$, which is again of the same
form as in the statement of the theorem.  For $R_j$s not of this form
there are no recursive calls, so nothing is added to $\mathcal
C$. Hence, by induction on the first argument to $\mathsf{adm}$, all
of the constraints added to $\mathcal C$ are of the form $\lop \Psi
\ol \phi, \Psi \ol \psi \rop$ for some type expression $\Psi$ and some
$\phi$s and $\psi$s, where the $\phi$s and $\psi$s are all pairwise
distinct from one another.

Each constraint of the form $\lop \Psi \ol \phi, \Psi \ol \psi \rop$
is top-unifiable and thus leads to a sequence of assignments of the
form $\lop \phi_i, \psi_i \rop$. Moreover, the fact that
$\tau_i\ol\beta\ol\gamma = \Sigma_i \ol\beta$ in Step (iii) ensures
that no $h$s appear in any $\zeta_{j,i} \ol g \ol h$, so the solved
constraints introduced by each recursive call can have as their
right-hand sides only $g$s introduced in the call from which they
spawned. It is not hard to see that the entire solved system resulting
from the original call must comprise the assignments $\lop g_{1,1},
f_1 \rop,...,\lop g_{1,n}, f_n \rop$ from the top-level call, as well
as the assignments $\lop g_{j_{i}+1,1}, g_{j_{i},1} \rop,...,\lop
g_{j_{i}+1,n}, g_{j_{i},n} \rop$, for $j_i = 0,...,m_i-1$ and $i =
1,...,n$, where $m_i$ is determined by the subtree of recursive calls
spawned by $f_i$.
%
Re-grouping this ``breadth-first'' collection of assignments
``depth-first'' by the trace of each $f_i$ for $i = 1,...,n$, we get a
solved system of the desired form.\looseness=-1
\end{proof}



\section{Examples}\label{sec:examples}

\begin{example}
For $t$ as in Example~\ref{ex:ex1}, the call
$~\mathsf{adm}\;t\;\;f\;\;\mathsf{Seq}\,\beta_1~$ results in the
sequence of calls:

\begin{figure}[H]
\centering
\scalebox{.85}{
\begin{tabular}{|l|l l l l|}
  \hline
  \emph{call 1} & $\mathsf{adm}$ & $t$ & $f$ &
  $\mathsf{Seq}\,\beta_1$\\ \hline 
  \emph{call 2.1} & $\mathsf{adm}$ &
  $\mathsf{pair}\;(\mathsf{const\,tt})\;(\mathsf{const\,2})$ & $h^1_1$
  & $\mathsf{Seq}\,\gamma^1_1$\\ \hline 
  \emph{call 2.2} & $\mathsf{adm}$ & $\mathsf{const\,5}$ & $h^1_2$ &
  $\mathsf{Seq}\,\gamma^1_2$\\ \hline
  \emph{call 2.1.1} & $\mathsf{adm}$ & $\mathsf{const\,tt}$ & $h^{2.1}_1$ &
  $\mathsf{Seq}\,\gamma^{2.1}_1$\\ \hline
  \emph{call 2.1.2} & $\mathsf{adm}$ & $\mathsf{const\,2}$ & $h^{2.1}_2$ &
  $\mathsf{Seq}\,\gamma^{2.1}_2$\\ \hline   
\end{tabular}}
\end{figure}
\noindent
The steps of $\mathsf{adm}$ corresponding to these call are given in
the table below, with the most important components of these steps
listed explicitly:

\begin{figure}[H]
\centering
\scalebox{.85}{
\begin{tabular}{|p{0.3cm}|p{2.4cm}|p{3cm}|p{2.3cm}|p{3.3cm}|p{2.8cm}|}
\hline
\thead{step} & \thead{{matching}} & \thead{$\ol\tau$} & \thead{$\ol
  R$} & \thead{$\ol{\zeta}$} & \thead{{constraints}}\\
\thead{no.} & \thead{{problems}} & \thead{ } &\thead{ } & \thead{ } &
  \thead{{added to $C$}}\\\hline\hline 
  \emph{1}
& $\beta_1 \equiv \gamma^1_1 \times \gamma^1_2$
& $\tau_1 \beta_1 \gamma^1_1 \gamma^1_2 = \gamma^1_1$ \newline $\tau_2
  \beta_1 \gamma^1_1 \gamma^1_2 = \gamma^1_2$ 
& $R_1 = \mathsf{Seq}\,\gamma^1_1$ \newline
  $R_2 = \mathsf{Seq}\,\gamma^1_2$
& $\zeta_{1,1}\beta_1\gamma^1_1\gamma^1_2 = \gamma^1_1$ \newline 
  $\zeta_{2,1}\beta_1\gamma^1_1\gamma^1_2 = \gamma^1_2$
  & $\lop g^1_1, f \rop$ \newline $\lop h^1_1 \times h^1_2, g^1_1
  \rop$ \\\hline
  \emph{2.1}
& $\gamma^1_1 \equiv \gamma^{2.1}_1 \times \gamma^{2.1}_2$
& $\tau_1 \gamma^1_1 \gamma^{2.1}_1 \gamma^{2.1}_2 =
  \gamma^{2.1}_1$ \newline $\tau_2 \gamma^1_1 \gamma^{2.1}_1
  \gamma^{2.1}_2 = \gamma^{2.1}_2$ 
& $R_1 = \mathsf{Seq}\,\gamma^{2.1}_1$ \newline
  $R_2 = \mathsf{Seq}\,\gamma^{2.1}_2$
& $\zeta_{1,1}\gamma^1_1\gamma^{2.1}_1\gamma^{2.1}_2 =
  \gamma^{2.1}_1$ \newline
  $\zeta_{2,1}\gamma^1_1\gamma^{2.1}_1\gamma^{2.1}_2 = 
  \gamma^{2.1}_2$ 
  & $\lop g^{2.1}_1, h^1_1 \rop$ \newline $\lop h^{2.1}_1 \times
  h^{2.1}_2, g^{2.1}_1 \rop$ 
  \\\hline
  \emph{2.2}
& $\gamma^2_1 \equiv \gamma^{2.2}_1$
& $\tau_1 \gamma^1_2 \gamma^{2.2}_1 = \gamma^1_2$
& $R_1 = \gamma^1_2$ 
& 
& $\lop g^{2.2}_1, h^1_2 \rop$ 
  \\\hline
  \emph{2.1.1}
& $\gamma^{2.1}_1 \equiv \gamma^{2.1.1}_1$
& $\tau_1 \gamma^{2.1}_1 \gamma^{2.1.1}_1 = \gamma^{2.1}_1$
& $R_1 = \gamma^{2.1}_1$ 
& 
& $\lop g^{2.1.1}_1, h^{2.1}_1 \rop$ 
  \\\hline
  \emph{2.1.2}
& $\gamma^{2.1}_2 \equiv \gamma^{2.1.2}_1$
& $\tau_1 \gamma^{2.1}_2 \gamma^{2.1.2}_1 = \gamma^{2.1}_2$
& $R_1 = \gamma^{2.1}_2$ 
& 
& $\lop g^{2.1.2}_1, h^{2.1}_2 \rop$ 
  \\\hline
\end{tabular}}
\end{figure}
\noindent
Since the solution to the generated set of constraints imposes the
requirement that $f = (g^{2.1.1}_1 \times g^{1.2.1}_1) \times
g^{2.2}_1$, we conclude that the most general functions mappable over
$t$ relative to the specification $\mathsf{Seq}\,\beta_1$ are those of
the form $f = (f_1 \times f_2) \times f_3$ for some types $X_1$,
$X_2$, and $X_3$ and functions $f_1 : \mathsf{Bool} \to X_1$, $f_2 :
\mathsf{Int} \to X_2$, and $f_3 : \mathsf{Int} \to X_3$.  This is
precisely the result obtained informally in Example~\ref{ex:ex1}.
\end{example}

\vspace*{0.05in}

\begin{example}\label{ex:u}
For $\mathsf{G}$ and $t$ as in Example~\ref{ex:ex2} and $f :
\mathsf{List}\,\mathbb{N} \times \mathbb{N} \to X$ the call
$~\mathsf{adm}\;t\;\;f\;\;\mathsf{G}\,\beta_1~$ results in the
sequence of calls:

\begin{figure}[H]
\centering
\scalebox{.85}{
\begin{tabular}{|l|l l l l|}
  \hline
  \emph{call 1} & $\mathsf{adm}$ & $t$ & $f$ & $\mathsf{G}\beta_1$\\ \hline
  \emph{call 2} & $\mathsf{adm}$ & $t_2$ & $\mathsf{G}h^1_1 \times
  \mathsf{G}(h^1_2 \times h^1_2)$  & $\mathsf{G}(\mathsf{G}\gamma^1_1
  \times \mathsf{G}(\gamma^1_2  \times \gamma^1_2))$\\ \hline
  \emph{call 3} & $\mathsf{adm}$ & $t_3$ & $(\mathsf{G}g^2_1,
  \mathsf{G}(g^2_2 \times g^2_2))$ & $\mathsf{G}\gamma^1_1 \times
  \mathsf{G}(\gamma^1_2 \times \gamma^1_2)$\\ \hline
  \emph{call 4.1} & $\mathsf{adm}$ &
  ${\mathsf{inj}}\;({\mathsf{cons}}\;2\;{\mathsf{nil}})$
  & $g^3_1$ & $\mathsf{G}\gamma^2_1$\\ \hline
  \emph{call 4.2} & $\mathsf{adm}$ & ${\mathsf{pairing}}\;
  ({\mathsf{inj}\,2})\;{\mathsf{const}}$ & $g^3_2 \times 
  g_2^3$ & $\mathsf{G}(\gamma^2_2 \times \gamma^2_2)$\\ \hline
  \emph{call 4.2.1} & $\mathsf{adm}$ & $\mathsf{inj}\,2$ &
  $g_1^{4.2}$
  & $\mathsf{G}\gamma^2_2$\\ \hline
  \emph{call 4.2.2} & $\mathsf{adm}$ & $\mathsf{const}$  &
  $g_1^{4.2}$
  & $\mathsf{G}\gamma^2_2$\\ \hline
\end{tabular}}
\end{figure}
\noindent
where
\[\begin{array}{lll}
t & = & \mathsf{projpair} \;(\;{\mathsf{inj}}\;
(\,{\mathsf{inj}}\;({\mathsf{cons}}\;2\;{\mathsf{nil}}),\;
{\mathsf{pairing}}\; ({\mathsf{inj}\,2})\;{\mathsf{const}}\;)\;)\\
t_2 & = & {\mathsf{inj}}\;
(\;{\mathsf{inj}}\;({\mathsf{cons}}\;2\;{\mathsf{nil}}),\;
{\mathsf{pairing}}\; ({\mathsf{inj}\,2})\;{\mathsf{const}}\;)\\
t_3 & = & (\;{\mathsf{inj}}\;({\mathsf{cons}}\;2\;{\mathsf{
    nil}}),\; {\mathsf{pairing}}\;
({\mathsf{inj}\,2})\;{\mathsf{const}}\;)
\end{array}\]
\noindent
The steps of $\mathsf{adm}$ corresponding to these call are given in
Table~\ref{fig:u}, with the most important components of these steps
listed explicitly. Since the solution to the generated set of
constraints imposes the requirement that $f = g^{4.1}_1
\times \id_\mathbb{N}$, we conclude that the most general functions
mappable over $t$ relative to the specification $\mathsf{G}\,\beta_1$
are those of the form $f = f' \times \id_\mathbb{N}$ for some type $X$
and some function $f' : \mathsf{List}\,\mathbb{N} \to X$. This is
precisely the result obtained intuitively in
Example~\ref{ex:ex2}.\looseness=-1

\begin{sidewaystable}
\begin{minipage}{0.5\textwidth}
\centering
\scalebox{.85}{
\begin{tabular}{|p{1cm}|p{3.5cm}|p{4.5cm}|p{4cm}|p{4.5cm}|p{5.7cm}|}
\hline
\thead{\emph{call}} &
  \thead{\emph{matching}}&
  \thead{$\ol\tau$} & \thead{$\ol R$} & \thead{$\ol\zeta$} &
  \thead{\emph{constraints}}\\
\thead{\emph{no.}} & \thead{\emph{problems}} &\thead{ }
&\thead{ } &\thead{ } &\thead{\emph{added to $C$}}\\\hline\hline
1
     & $\beta_1 \equiv \gamma^1_1 \times \gamma^1_1$
     & $\tau_1 \beta_1 \gamma^1_1\gamma^1_2 = \gamma^1_1$ \newline
       $\tau_2 \beta_1 \gamma^1_1\gamma^1_2 = \gamma^1_2$
     & $R_1 = \mathsf{G}\,(\mathsf{G}\gamma^1_1 \times
       \mathsf{G}(\gamma^1_2 \times \gamma^1_2))$ 
     & $\zeta_{1,1} \beta_1\gamma^1_1\gamma^1_2 = \mathsf{G}\gamma^1_1
       \times \mathsf{G}(\gamma^1_2 \times \gamma^1_2)$ 
     & $\lop g_1^1,f\rop$ \newline $\lop h_1^1 \times h_2^1, g_1^1 \rop$
     \\\hline
     2
     & $\mathsf{G}\gamma^1_1 \times \mathsf{G}(\gamma^1_2 \times \gamma^1_2)
     \equiv \gamma_1^2$ 
     & $\tau_1\gamma^1_1\gamma^1_2\gamma_1^2 = \mathsf{G}\gamma^1_1 \times
     \mathsf{G}(\gamma^1_2 \times \gamma^1_2)$ 
     & $R_1 = \mathsf{G}\gamma^1_1 \times
     \mathsf{G}(\gamma^1_2 \times \gamma^1_2)$
     & $\zeta_{1,1}\gamma^1_1\gamma^1_2\gamma_1^2 = \mathsf{G}\gamma^1_1$ \newline
       $\zeta_{1,2}\gamma^1_1\gamma^1_2\gamma_1^2 = \mathsf{G}(\gamma^1_2
     \times \gamma^1_2)$ 
     & $\lop \mathsf{G}g_1^2 \times \mathsf{G}(g_2^2 \times
     g_2^2), \mathsf{G}h_1^1 \times \mathsf{G}(h^1_2
     \times h^1_2) \rop$\\ \hline
     3
     &
     &
     &
     & $\zeta_1\gamma^2_1\gamma^2_2 = \gamma^2_1$ \newline
       $\zeta_2\gamma^2_1\gamma^2_2 = \gamma^2_2 \times \gamma^2_2$
     & $ \lop \mathsf{G}g_1^3, \mathsf{G}g_1^2 \rop$ \newline
       $ \lop \mathsf{G}(g_2^3 \times g_2^3), \mathsf{G}(g_2^2 \times
     g_2^2) \rop$\\ 
     \hline
     4.1
     & $\gamma^2_1 \equiv \gamma_1^{4.1}$
     & $\tau_1\gamma^2_1\gamma_1^{4.1} = \gamma_1^2$
     & $R_1 = \gamma^2_1$
     & 
     & $\lop g_1^{4.1}, g_1^3 \rop$\\ 
     \hline
     4.2
     & $\gamma^2_2 \times \gamma^2_2 \equiv \gamma_1^{4.2} \times \gamma_2^{4.2}$
     & $\tau_1\gamma^2_2\gamma_1^{4.2}\gamma_2^{4.2} = \gamma^2_2$ \newline
     $\tau_2\gamma^2_2\gamma_1^{4.2}\gamma_2^{4.2} = \gamma^2_2$
     & $R_1 = \mathsf{G}\gamma^2_2$ \newline
       $R_2 = \mathsf{G}\gamma^2_2$ 
     & $\zeta_{1,1}\gamma^2_1\gamma_1^{4.2}\gamma_2^{4.2} = \gamma^2_2$ \newline
       $\zeta_{2,1}\gamma^2_1\gamma_1^{4.2}\gamma_2^{4.2} = \gamma^2_2$
     & $\lop g_1^{4.2} \times g_1^{4.2}, g_2^3 \times g_2^3 \rop$
     \\
     \hline
     4.2.1
     & $\gamma^2_2 \equiv \gamma_1^{4.2.1}$
     & $\tau_1\gamma^2_2\gamma_1^{4.2.1} = \gamma^2_2$
     & $R_1 = \gamma^2_2$
     & 
     & $\lop g_1^{4.2.1}, g_1^{4.2} \rop$ 
     \\
     \hline
     4.2.2
     & $\gamma^2_2 \equiv \mathbb{N}$
     & 
     & $R_1 = 1$
     & 
     & $\lop g_1^{4.2.2},g_1^{4.2} \rop$ \newline $\lop
     \id_{\mathbb{N}},g_1^{4.2.2} \rop$  
     \\
     \hline
\end{tabular}}
\caption{Calls for Example~\ref{ex:u}}
\label{fig:u}
\end{minipage}

\vspace*{0.5in}

\begin{minipage}{0.5\textwidth}
\centering
\scalebox{.85}{
\begin{tabular}{|p{1cm}|p{3.5cm}|p{4.5cm}|p{4cm}|p{4.5cm}|p{5.7cm}|}
\hline
\thead{\emph{call}} &
  \thead{\emph{matching}}&
  \thead{$\ol\tau$} & \thead{$\ol R$} & \thead{$\ol\zeta$} &
  \thead{\emph{constraints}}\\
\thead{\emph{no.}} & \thead{\emph{problems}} &\thead{ }
&\thead{ } &\thead{ } &\thead{\emph{added to $C$}}\\\hline\hline
1
     & $\beta_1 \equiv \gamma^1_1 \times \gamma^1_1$
     & $\tau_1 \beta_1 \gamma^1_1\gamma^1_2 = \gamma^1_1$ \newline
       $\tau_2 \beta_1 \gamma^1_1\gamma^1_2 = \gamma^1_2$
     & $R_1 = \mathsf{G}\,(\mathsf{G}\gamma^1_1 \times
       \mathsf{G}(\gamma^1_2 \times \gamma^1_2))$ 
     & $\zeta_{1,1} \beta_1\gamma^1_1\gamma^1_2 = \mathsf{G}\gamma^1_1
       \times \mathsf{G}(\gamma^1_2 \times \gamma^1_2)$ 
     & $\lop g_1^1,f\rop$ \newline $\lop h_1^1 \times h_2^1, g_1^1 \rop$
     \\\hline
     2
     & $\mathsf{G}\gamma^1_1 \times \mathsf{G}(\gamma^1_2 \times \gamma^1_2)
     \equiv \gamma_1^2$ 
     & $\tau_1\gamma^1_1\gamma^1_2\gamma_1^2 = \mathsf{G}\gamma^1_1 \times
     \mathsf{G}(\gamma^1_2 \times \gamma^1_2)$ 
     & $R_1 = \mathsf{G}\gamma^1_1 \times
     \mathsf{G}(\gamma^1_2 \times \gamma^1_2)$
     & $\zeta_{1,1}\gamma^1_1\gamma^1_2\gamma_1^2 = \mathsf{G}\gamma^1_1$ \newline
       $\zeta_{1,2}\gamma^1_1\gamma^1_2\gamma_1^2 = \mathsf{G}(\gamma^1_2
     \times \gamma^1_2)$ 
     & $\lop \mathsf{G}g_1^2 \times \mathsf{G}(g_2^2 \times
     g_2^2), \mathsf{G}h_1^1 \times \mathsf{G}(h^1_2
     \times h^1_2) \rop$\\ \hline
     3
     &
     &
     &
     & $\zeta_1\gamma^2_1\gamma^2_2 = \gamma^2_1$ \newline
       $\zeta_2\gamma^2_1\gamma^2_2 = \gamma^2_2 \times \gamma^2_2$
     & $ \lop \mathsf{G}g_1^3, \mathsf{G}g_1^2 \rop$ \newline
       $ \lop \mathsf{G}(g_2^3 \times g_2^3), \mathsf{G}(g_2^2 \times
     g_2^2) \rop$\\ 
     \hline
     4.1
     & $\gamma^2_1 \equiv \mathsf{List}\,\gamma_1^{4.1}$
     & $\tau_1\gamma^2_1\gamma_1^{4.1} = \gamma_1^{4.1}$
     & $R_1 = \mathsf{List}\,(\mathsf{G}\gamma_1^{4.1})$
     & $\zeta_{1,1}\gamma^2_1\gamma_1^{4.1} = \mathsf{G}\gamma_1^{4.1}$
     & $\lop g_1^{4.1}, g_1^3 \rop$ \newline
       $\lop \mathsf{List}\,h_1^{4.1}, g_1^{4.1} \rop$ \\ 
     \hline
     4.2
     & $\gamma^2_2 \times \gamma^2_2 \equiv \gamma_1^{4.2} \times \gamma_2^{4.2}$
     & $\tau_1\gamma^2_2\gamma_1^{4.2}\gamma_2^{4.2} = \gamma^2_2$ \newline
     $\tau_2\gamma^2_2\gamma_1^{4.2}\gamma_2^{4.2} = \gamma^2_2$
     & $R_1 = \mathsf{G}\gamma^2_2$ \newline
       $R_2 = \mathsf{G}\gamma^2_2$ 
     & $\zeta_{1,1}\gamma^2_1\gamma_1^{4.2}\gamma_2^{4.2} = \gamma^2_2$ \newline
       $\zeta_{2,1}\gamma^2_1\gamma_1^{4.2}\gamma_2^{4.2} = \gamma^2_2$
     & $\lop g_1^{4.2} \times g_1^{4.2}, g_2^3 \times g_2^3 \rop$
     \\
     \hline
     4.1.1
     & $\mathsf{G}\gamma_1^{4.1} \equiv\gamma_1^{4.1.1}$
     & $\tau_1\gamma_1^{4.1}\gamma_1^{4.1.1} = \mathsf{G}\gamma_1^{4.1}$ 
     & $R_1 = \mathsf{G}\gamma_1^{4.1}$ \newline $R_2 =
     \mathsf{List}(\mathsf{G}\gamma_1^{4.1})$ 
     & $\zeta_{1,1}\gamma_1^{4.1}\gamma_1^{4.1.1} =
     \gamma_1^{4.1}$ \newline
     $\zeta_{2,1}\gamma^{4.1}_1\gamma^{4.1.1}_1 = \mathsf{G}\gamma^{4.1}_1$
     & $\lop \mathsf{G}g_1^{4.1.1}, \mathsf{G}h_1^{4.1} \rop$
     \\
     \hline
     4.2.1
     & $\gamma^2_2 \equiv \gamma_1^{4.2.1}$
     & $\tau_1\gamma^2_2\gamma_1^{4.2.1} = \gamma^2_2$
     & $R_1 = \gamma^2_2$
     & 
     & $\lop g_1^{4.2.1},  g_1^{4.2} \rop$ 
     \\
     \hline
     4.2.2
     & $\gamma^2_2 \equiv \mathbb{N}$
     & 
     & $R_1 = 1$
     & 
     & $\lop g_1^{4.2.2},g_1^{4.2} \rop$ \newline $\lop
     \id_{\mathbb{N}},g_1^{4.2.2} \rop$  
     \\
     \hline
     4.1.1.1
     & $\gamma^{4.1}_1 \equiv \mathbb{N}$ 
     & 
     & $R_1 = 1$
     &
     & $\lop g_1^{4.1.1.1}, g_1^{4.1.1} \rop$ \newline
       $\lop \id_{\mathbb{N}}, g_1^{4.1.1.1} \rop$
     \\
     \hline
     4.1.1.2
     & $\mathsf{G}\gamma_1^{4.1} \equiv \gamma_1^{4.1.1.2}$
     & $\tau_1\gamma_1^{4.1}\gamma_1^{4.1.1.2} = \mathsf{G}\gamma_1^{4.1}$
     & $R_1 = 1$
     & 
     & $\lop \mathsf{G}g_1^{4.1.1.2}, \mathsf{G}g_1^{4.1.1} \rop$ 
     \\
     \hline
\end{tabular}}
\caption{Calls for Example~\ref{ex:t}}
\label{fig:t}
\end{minipage}
\end{sidewaystable}
\end{example}

\begin{example}\label{ex:t}
For $\mathsf{G}$ and $t$ as in Example~\ref{ex:ex3}
and $f : \mathsf{List}\,\mathbb{N} \times \mathbb{N} \to X$ we have
\[\begin{array}{lll}
K^{\mathsf{const}} & = & \mathbb{N}\\
K^{\mathsf{flat}}\,\alpha & = & \mathsf{List}\,\alpha\\
K^{\mathsf{inj}}\,\alpha & = & \alpha\\
K^{\mathsf{pairing}}\,\alpha_1\,\alpha_2 & = & \alpha_1 \times \alpha_2\\
K^{\mathsf{projpair}}\,\alpha_1\,\alpha_2 & = & \alpha_1 \times \alpha_2
\end{array}\]
The call $~\mathsf{adm}\;t\;\;f\;\;\mathsf{G}\,\beta_1~$ results in the
sequence of calls:

\begin{figure}[H]
\centering
\scalebox{.85}{
\begin{tabular}{|l|l l l l|}
  \hline
  \emph{call 1} & $\mathsf{adm}$ & $t$ & $f$ & $\mathsf{G}\beta_1$\\ \hline
  \emph{call 2} & $\mathsf{adm}$ & $t_2$ & $\mathsf{G}h^1_1 \times
  \mathsf{G}(h^1_2 \times h^1_2)$  & $\mathsf{G}(\mathsf{G}\gamma^1_1
  \times \mathsf{G}(\gamma^1_2  \times \gamma^1_2))$\\ \hline
  \emph{call 3} & $\mathsf{adm}$ & $t_3$ & $(\mathsf{G}g^2_1,
  \mathsf{G}(g^2_2 \times g^2_2))$ & $\mathsf{G}\gamma^1_1 \times
  \mathsf{G}(\gamma^1_2 \times \gamma^1_2)$\\ \hline
  \emph{call 4.1} & $\mathsf{adm}$ &
  ${\mathsf{flat}}\;({\mathsf{cons}}\;{\mathsf{const}}\;{\mathsf{nil}})$
  & $g^3_1$ & $\mathsf{G}\gamma^2_1$\\ \hline
  \emph{call 4.2} & $\mathsf{adm}$ & ${\mathsf{pairing}}\;
  ({\mathsf{inj}\,2})\;{\mathsf{const}}$ & $g^3_2 \times 
  g_2^3$ & $\mathsf{G}(\gamma^2_2 \times \gamma^2_2)$\\ \hline
  \emph{call 4.1.1} & $\mathsf{adm}$ &
  $\mathsf{cons}\,\mathsf{const}\,\mathsf{nil}$  & $\mathsf{G}h_1^{4.1}$ &
  $\mathsf{List}\,(\mathsf{G}\gamma_1^{4.1})$\\ \hline
  \emph{call 4.2.1} & $\mathsf{adm}$ & $\mathsf{inj}\,2$ &
  $g_1^{4.2}$
  & $\mathsf{G}\gamma^2_2$\\ \hline
  \emph{call 4.2.2} & $\mathsf{adm}$ & $\mathsf{const}$  &
  $g_1^{4.2}$
  & $\mathsf{G}\gamma^2_2$\\ \hline
  \emph{call 4.1.1.1} & $\mathsf{adm}$ & $\mathsf{const}$ &
  $g_1^{4.1.1}$ & $\mathsf{G}\,\gamma_1^{4.1}$\\ \hline
  \emph{call 4.1.1.2} & $\mathsf{adm}$ & $\mathsf{nil}$ &
  $\mathsf{G}g_1^{4.1.1}$ & $\mathsf{List}(\mathsf{G}\gamma_1^{4.1})$\\ \hline
\end{tabular}}
\end{figure}
\noindent
where
\[\begin{array}{lll}
t & = & \mathsf{projpair} \;(\;{\mathsf{inj}}\;
(\,{\mathsf{flat}}\;({\mathsf{cons}}\;{\mathsf{const}}\;{\mathsf{
    nil}}),\; {\mathsf{pairing}}\;
({\mathsf{inj}\,2})\;{\mathsf{const}}\;)\;)\\
t_2 & = & {\mathsf{inj}}\;
(\;{\mathsf{flat}}\;({\mathsf{cons}}\;{\mathsf{const}}\;{\mathsf{
    nil}}),\; {\mathsf{pairing}}\;
({\mathsf{inj}\,2})\;{\mathsf{const}}\;)\\
t_3 & = & (\;{\mathsf{flat}}\;({\mathsf{cons}}\;{\mathsf{const}}\;{\mathsf{
    nil}}),\; {\mathsf{pairing}}\;
({\mathsf{inj}\,2})\;{\mathsf{const}}\;)
\end{array}\]
\noindent
The steps of $\mathsf{adm}$ corresponding to these call are given in
Table~\ref{fig:t}, with the most important components of these steps
listed explicitly.  Since the solution to the generated set of
constraints imposes the requirement that $f =
map_{\mathsf{List}}\,\id_{\mathbb{N}} \; \times \; \id_\mathbb{N}$,
we conclude that the only function mappable over $t$ relative to the
specification $\mathsf{G}\,\beta_1$ is this $f$. This is precisely
the result obtained informally in Example~\ref{ex:ex3}.
\end{example}

\begin{example}\label{ex:list}
For $t$ as in Example~\ref{ex:ex4} the call
$~\mathsf{adm}\;t\;\;f\;\;\mathsf{List}\,\beta_1~$ results in the
sequence of calls:

\begin{figure}[H]
\centering
\scalebox{.85}{
\begin{tabular}{|l|l l l l|}
  \hline
\emph{call 1} & $\mathsf{adm}$ & $t$ & $f$ &
  $\mathsf{List}\,\beta_1$\\ \hline
\emph{call 2} & $\mathsf{adm}$ &
$\mathsf{cons}\,(\mathsf{cons\,3\,nil})\,\mathsf{nil})$ &
$g^1_1$ &  $\mathsf{List}\,\beta_1$\\ \hline
\emph{call 2.1} & $\mathsf{adm}$ &
$\mathsf{nil}$ &  $g^2_1$ &  $\mathsf{List}\,\beta_1$\\ \hline  
\end{tabular}}
\end{figure}
\noindent
The steps of $\mathsf{adm}$ corresponding to these call are given in
the table below, with the most important components of these steps
listed explicitly:

\begin{figure}[H]
\centering
\scalebox{.85}{
\begin{tabular}{|p{0.3cm}|p{1.5cm}|p{2.3cm}|p{2cm}|p{2.1cm}|p{0.5cm}|}
\hline
\thead{step} & \thead{{matching}} & \thead{$\ol\tau$} & \thead{$\ol
  R$} & \thead{$\ol{\zeta}$} & \thead{{constraints}}\\
\thead{no.} & \thead{{problems}} & \thead{ } &\thead{ } & \thead{ } &
  \thead{{added to $C$}}\\\hline\hline 
  \emph{1}
& $\beta_1 \equiv \gamma^1_1$
& $\tau_1 \beta_1 \gamma^1_1 = \beta_1$
& $R_1 = \beta_1$ \newline
  $R_2 = \mathsf{List}\,\beta_1$
& $\zeta_{2,1}\beta_1\gamma^1_1 = \beta_1$
  & $\lop g^1_1, f \rop$
  \\\hline
  \emph{2}
& $\beta_1 \equiv \gamma^2_1$
& $\tau_1 \beta_1 \gamma^2_1 = \beta_1$
& $R_1 = \beta_1$ \newline
  $R_2 = \mathsf{List}\,\beta_1$
& $\zeta_{2,1}\beta_1\gamma^2_1 = \beta_1$
  & $\lop g^2_1, g^1_1 \rop$ 
  \\\hline
  \emph{2.1}
& $\beta_1 \equiv \gamma^{2.1}_1$
& $\tau_1 \beta_1 \gamma^{2.1}_1 = \beta_1$
& $R_1 = 1$ 
& 
& $\lop g^{2.1}_1, g^2_1 \rop$ 
  \\\hline
\end{tabular}}
\end{figure}
\noindent
Since the solution to the generated set of constraints imposes the
requirement that $f = g^{2.1}_1$, we conclude that any function $f :
\mathsf{List}\,\mathbb{N} \to X$ (for some type $X$) is mappable over
$t$ relative to the specification $\mathsf{List}\,\beta_1$.
\end{example}

\begin{example}\label{ex:ex5-again}
For $t$ as in Example~\ref{ex:ex5} the call
$~\mathsf{adm}\;t\;\;f\;\;\mathsf{List\,(List\,\beta_1)}~$ results in
the following sequence of calls:\looseness=-1

\vspace*{0.1in}

\begin{figure}[H]
\centering
\scalebox{.85}{
\begin{tabular}{|l|l l l l|}
  \hline
\emph{call 1} & $\mathsf{adm}$ & $t$ & $f$ &
  $\mathsf{List}\,\beta_1$\\ \hline
\emph{call 2.1} & $\mathsf{adm}$ &
$\mathsf{cons\,1}\,(\mathsf{cons\,2\,nil})$ & $g^1_1$ &
  $\mathsf{List}\,\beta_1$\\ \hline
\emph{call 2.2} & $\mathsf{adm}$ &
$\mathsf{cons}\,(\mathsf{cons\,3\,nil})\,\mathsf{nil})$ &
  $\mathsf{List}\,g^1_1$ &
  $\mathsf{List}\,(\mathsf{List}\,\beta_1)$\\ \hline  
\emph{call 2.1.1} & $\mathsf{adm}$ &
$\mathsf{cons\,2\,nil}$ & $g^{2.1}_1$ &
  $\mathsf{List}\,\beta_1$\\ \hline
\emph{call 2.2.1} & $\mathsf{adm}$ &
$\mathsf{cons\,3\,nil}$ & $g^{2.2}_1$ &
  $\mathsf{List}\,\beta_1$\\ \hline
\emph{call 2.2.2} & $\mathsf{adm}$ &
$\mathsf{nil}$ & $\mathsf{List}\,g^{2.2}_1$ &
  $\mathsf{List}\,(\mathsf{List}\,\beta_1)$\\ \hline
\emph{call 2.1.1.1} & $\mathsf{adm}$ &
$\mathsf{nil}$ & $g^{2.1.1}_1$ &
  $\mathsf{List}\,\beta_1$\\ \hline
\emph{call 2.2.1.1} & $\mathsf{adm}$ &
$\mathsf{nil}$ & $g^{2.2.1}_1$ &
  $\mathsf{List}\,\beta_1$\\ \hline
\end{tabular}}
\end{figure}

\vspace*{0.1in}

\noindent
The steps of $\mathsf{adm}$ corresponding to these calls are given in
the table below, with the most important components of these steps
listed explicitly:

\vspace*{0.1in}

\begin{figure}[H]
\centering
\scalebox{.85}{
\begin{tabular}{|p{1cm}|p{2.4cm}|p{2.9cm}|p{2.8cm}|p{3.5cm}|p{2.8cm}|}
\hline
\thead{step} & \thead{{matching}} & \thead{$\ol\tau$} & \thead{$\ol
  R$} & \thead{$\ol{\zeta}$} & \thead{{constraints}}\\
\thead{no.} & \thead{{problems}} & \thead{ } &\thead{ } & \thead{ } &
  \thead{{added to $C$}}\\\hline\hline 
  \emph{1}
& $\mathsf{List}\,\beta_1 \equiv \gamma^1_1$
& $\tau_1 \beta_1 \gamma^1_1 = \mathsf{List}\,\beta_1$
& $R_1 = \mathsf{List}\,\beta_1$ \newline
  $R_2 = \mathsf{List}\,(\mathsf{List}\,\beta_1)$
& $\zeta_{1,1}\beta_1\gamma^1_1 = \beta_1$ \newline
  $\zeta_{2,1}\beta_1\gamma^1_1 = \mathsf{List}\,\beta_1$
& $\lop \mathsf{List}\,g^1_1, f \rop$
  \\\hline
  \emph{2.1}
& $\beta_1 \equiv \gamma^{2.1}_1$
& $\tau_1 \beta_1 \gamma^{2.1}_1 = \beta_1$
& $R_1 = \beta_1$ \newline
  $R_2 = \mathsf{List}\,\beta_1$
& $\zeta_{2,2}\beta_1\gamma^{2.1}_1 = \beta_1$
& $\lop g^{2.1}_1, g^1_1 \rop$
  \\\hline
  \emph{2.2}
& $\mathsf{List}\,\beta_1 \equiv \gamma^{2.2}_1$
& $\tau_1 \beta_1 \gamma^{2.2}_1 = \mathsf{List}\,\beta_1$
& $R_1 = \mathsf{List}\,\beta_1$ \newline
  $R_2 = \mathsf{List}\,(\mathsf{List}\,\beta_1)$
& $\zeta_{1,1}\beta_1\gamma^{2.2}_1 = \beta_1$ \newline
  $\zeta_{2,1}\beta_1\gamma^{2.2}_1 = \mathsf{List}\,\beta_1$
& $\lop \mathsf{List}\, g^{2.2}_1, \mathsf{List}\,g^1_1 \rop$
\\\hline
  \emph{2.1.1}
& $\beta_1 \equiv \gamma^{2.1.1}_1$
& $\tau_1 \beta_1 \gamma^{2.1.1}_1 = \beta_1$
& $R_1 = \beta_1$ \newline
  $R_2 = \mathsf{List}\,\beta_1$
& $\zeta_{2,2}\beta_1\gamma^{2.1.1}_1 = \beta_1$
  & $\lop g^{2.1.1}_1, g^{2.1}_1 \rop$
  \\\hline
  \emph{2.2.1}
& $\beta_1 \equiv \gamma^{2.2.1}_1$
& $\tau_1 \beta_1 \gamma^{2.2.1}_1 = \beta_1$
& $R_1 = \beta_1$ \newline
  $R_2 = \mathsf{List}\,\beta_1$
& $\zeta_{2,2}\beta_1\gamma^{2.2.1}_1 = \beta_1$
& $\lop g^{2.2.1}_1, g^{2.2}_1 \rop$
  \\\hline
  \emph{2.2.2}
& $\mathsf{List}\,\beta_1 \equiv \gamma^{2.2.2}_1$
& $\tau_1 \beta_1 \gamma^{2.2.2}_1 = \mathsf{List}\,\beta_1$ 
& $R_1 = 1$
& 
& $\lop \mathsf{List}\,g^{2.2.2}_1, \mathsf{List}\,g^{2.2}_1 \rop$
  \\\hline
  \emph{2.1.1.1}
& $\beta_1 \equiv \gamma^{2.1.1.1}_1$
& $\tau_1 \beta_1 \gamma^{2.1.1.1}_1 = \beta_1$
& $R_1 = 1$
& 
& $\lop g^{2.1.1.1}_1, g^{2.1.1}_1 \rop$
\\\hline
  \emph{2.2.1.1}
& $\beta_1 \equiv \gamma^{2.2.1.1}_1$
& $\tau_1 \beta_1 \gamma^{2.2.1.1}_1 = \beta_1$
& $R_1 = 1$
& 
& $\lop g^{2.2.1.1}_1, g^{2.2.1}_1 \rop$
\\\hline
\end{tabular}}
\end{figure}
\noindent
Since the solution to the generated set of constraints imposes the
requirement that $f = \mathsf{List}\,g^{2.2.1.1}_1$, we conclude that
the most general functions mappable over $t$ relative to the
specification $\mathsf{List}\,(\mathsf{List}\,\beta_1)$ are those of
the form $f = \mathsf{map_{List}}\,f'$ for some type $X$ and function
$f' : \mathbb{N} \to X$.
\end{example}

\section{Conclusion and Future
  Directions}\label{sec:conclusion}

The work reported here is part of a larger effort to develop a single,
unified categorical theory of data types. In particular, it can be
seen as a first step toward a properly functorial initial algebra
semantics for GADTs that specializes to the standard functorial
initial algebra semantics for nested types (which itself subsumes the
standard such semantics for ADTs) whenever the GADTs in question is a
nested type (or ADT).\looseness=-1

Categorical semantics of GADTs have been studied in \cite{hf11} and
\cite{jg08}. Importantly, both of these works interpret a GADT as a
fixpoint of a higher-order endofunctor $\funccat {U} \Set \to \funccat
{U} \Set$, where the category $U$ is {\em discrete}. As discussed in
Section~\ref{sec:introduction}, this destroys one of the main benefits
of interpreting a data type $\mathsf{D}$ as a fixpoint $\mu
F_\mathsf{D}$ of a higher-order endofunctor $F_\mathsf{D}$, namely the
existence of a non-trivial map function. Indeed, the action on
morphisms of $\mu F_\mathsf{D}$ should interpret the map function
$\mathsf{map_D}$ standardly associated with $\mathsf{D}$. But in the
discrete settings of \cite{hf11} and \cite{jg08}, the resulting
endofunctor $\mu F_\mathsf{D} : U \to \Set$ has very little to say
about the interpretation of $\mathsf{map_{D}}$, since its functorial
action need only specify the result of applying $\mathsf{map_{D}}$ to
a function $\mathsf{f : A \to B}$ when $\mathsf{B}$ is $\mathsf{A}$
and $\mathsf{f}$ is the identity function on $\mathsf{A}$. In
addition,~\cite{hf11} cannot handle truly nested data types such as
$\mathsf{Bush}$ or the GADT $\mathsf{G}$ from Example~\ref{ex:ex2}.
The resulting discrete initial algebra semantics for GADTs thus do not
recover the usual functorial initial algebra semantics of nested types
(including ADTs and truly nested types) when instantiated to these
special classes of GADTs.

In~\cite{fio12} an attempt is made to salvage the method
from~\cite{hf11} while taking the aforementioned issues into
account. The overall idea is to relax the discreteness requirement on
the category $U$, and to replace dependent products and sums in the
development of~\cite{hf11} with left and right Kan extensions,
respectively. But then the domain of $\mu F_{\mathsf{D}}$ must be the
category of all interpretations of types and all morphisms between
them, which in turn leads to the inclusion of unwanted junk elements
obtained by map closure, as already described in
Section~\ref{sec:introduction} of~\cite{jp19}. So this solution also
fails to bring us closer to a semantics of the kind we are aiming for.

{\em Containers}~\cite{aag03,aag05} provide an entirely different
approach to describing the functorial action of an ADT or nested type.
In this approach an element of such a type is described first by its
structure, and then by the data that structure contains. That is, a
ADT or nested type $\type D$ is seen as comprising a set
$S_{\mathsf{D}}$ of {\em shapes} and, for each shape $s\in
S_{\mathsf{D}}$, a set $P_{\,\mathsf{D},s}$ of {\em positions} in
$s$. If $\mathsf{A}$ is a type, then an element of $\type D\,\type A$
consists of a choice of a shape $s$ and a labeling of each of position
in $s$ by elements of $\type A$. Thus, if $A$ interprets $\type A$,
then $\type D\,\type A$ is interpreted as a labeling $\sum_{s\in
  S_{\type D}}(P_{\,\type D,s} \to A)$. The interpretation $D$ for
$\mathsf{D}$ simply abstracts this interpretation over $\mathsf{D}$'s
input type, and, for any morphism $f : A \to B$, the functorial action
$D\,f : \sum_{s\in S_{\type D}}(P_{\,\type D,s} \to A) \,\to\, \sum_{s\in
  S_{\type D}}(P_{\,\type D,s} \to B)$ is obtained by post-composition.
This functorial action does indeed interpret $\mathsf{map_{D}}$: given
a shape and a labeling of its position by elements of $\type A$, we
get automatically a data structure of the same shape whose positions
are labeled by elements of $\type B$ as soon as we have a function
$\mathsf{f : A \to B}$ to translate elements of $\type A$ to elements
of $\type B$.\looseness=-1

GADTs that go beyond ADTs and nested types have been studied from the
container point of view as {\em indexed containers}, both in
\cite{agh+15} and again in \cite{hf11}. The authors of \cite{agh+15}
propose encoding strictly positive indexed data types in terms of some
syntactic combinators they consider ``categorically
inspired''. However, as far as we understand their claim, map
functions and their interpretations as a functorial actions are not
worked out for indexed containers. The encoding in~\cite{agh+15}
nevertheless remains essential to understanding GADTs and other
inductive families as ``structures containing data''. With respect to
it, our algorithm can be understood as determining how ``containery''
a GADT $\mathsf{D}$ written in, say, Haskell or Agda is.  Indeed,
given a term $t$ whose type is an instance of $\mathsf{D}$, our
algorithm can determine $t$'s shape and positions, so there is no
longer any need to guess or otherwise divine them. Significantly,
there appears to be no general technique for determining the shapes
and positions of the elements of a data type just from the type's
programming language definition, and the ability to determine
appropriate shapes and position sets usually comes only with a deep
understanding of, and extensive experience with, the data structures
at play.\looseness=-1

We do not know of any other careful study of the functorial action of
type-indexed strictly positive inductive families. The work reported
here is the result of such a study for a specific class of such types,
namely the GADTs described in Equations~(\ref{eq:gadts}) and
(\ref{eq:data-constr-types}). Our algorithm defines map functions for
GADTs that coincide with the usual ones for GADTs that are ADTs and
nested types. The map functions computed by our algorithm will guide
our ongoing efforts to give functorial initial algebra semantics for
GADTs that subsume the usual ones for ADTs and nested types as
fixpoints of higher-order endofunctors.
\vspace*{0.1in}

\noindent
{\bf Acknowledgments} This research was supported in part by NSF award
CCR-1906388. It was performed while visiting Aarhus University's Logic
and Semantics group, which provided additional support via Villum
Investigator grant no.~25804, Center for Basic Research in Program
Verification.

\end{document}